\documentclass[aps,showpacs,pre,floatfix]{revtex4}
\usepackage{epsfig}
\usepackage{subfigure}
\usepackage[bahasa]{babel}
\begin{document}

\title{Analisis Kinerja Sistem Cluster Terhadapa Aplikasi Simulasi
Dinamika Molekular NAMD Memanfaatkan Pustaka CHARM++}
\author{A.B. Mutiara}
\affiliation{Magister Program on Information System, Gunadarma University,
\\Jl. Margonda Raya No. 100, Depok 16424, Indonesia\\
E-mail:amutiara@staff.gunadarma.ac.id}

\begin{abstract}

Tingkat kompleksitas dari program simulasi dinamika molekular
membutuhkan mesin pemroses dengan kemampuan yang sangat besar.
Mesin-mesin paralel terbukti memiliki potensi untuk menjawab
tantangan komputasi ini. Untuk memanfaatkan potensi ini secara
maksimal, diperlukan suatu program paralel dengan tingkat
efisiensi, efektifitas, skalabilitas, dan ekstensibilitas yang
maksimal pula.  Program NAMD yang dibahas pada penulisan ini
dianggap mampu untuk memenuhi semua kriteria yang diinginkan.
Program ini dirancang dengan mengimplementasikan pustaka Charm++
untuk pembagian tugas perhitungan secara paralel. NAMD memiliki
sistem automatic load balancing secara periodik yang cerdas,
sehingga dapat memaksimalkan penggunaan kemampuan mesin yang
tersedia. Program ini juga dirancang secara modular, sehingga
dapat dimodifikasi dan ditambah dengan sangat mudah. NAMD
menggunakan banyak kombinasi algoritma perhitungan dan
tehnik-tehnik numerik lainnya dalam melakukan tugasnya. NAMD 2.5
mengimplementasikan semua tehnik dan persamaan perhitungan yang
digunakan dalam dunia simulasi dinamika molekular saat ini. NAMD
dapat berjalan diatas berbagai mesin paralel termasuk arsitektur
cluster, dengan hasil speedup yang mengejutkan. Tulisan ini akan
menjelaskan dan membuktikan kemampuan NAMD secara paralel diatas
lima buah mesin cluster. Penulisan ini juga akan memaparkan
kinerja NAMD pada beberapa.

\textit{Kata kunci: Atom, Charm++, Cluster, Linux, Molekular,
NAMD, Paralel, Simulasi, VMD.}
\end{abstract}
\maketitle

\section{Pendahuluan}

Pesatnya perkembangan dunia teknologi informasi dewasa ini
merupakan suatu fenomena tersendiri. Berbagai terobosan dan
pembaruan dilakukan setiap menitnya, menjadikan dunia komputer
semakin cepat dan cepat saja. Kemajuan yang dicapai pada berbagai
bidang lain pun tidak terlepas dari bantuan komputer, apakah itu
dalam bidang komputasi, otomatisasi, ataupun simulasi, yang
kesemuanya berujung pada pemecahan masalah.

Ironisnya, pemecahan masalah berteknologi tinggi ini biasanya
diikuti pula dengan harga yang sangat tinggi. Harga mainframe dan
super komputer sudah tentu hanya dapat dijangkau oleh kalangan
tertentu saja. Apakah tidak ada hal yang dapat dilakukan? Haruskah
suatu penelitian menjadi terbengkalai karena tidak adanya dukungan
komputasi yang memadai?

Sistem cluster menjawab tantangan ini dengan menyediakan performa
perhitungan tingkat tinggi, namun dengan harga yang relatif
rendah. Bukan itu saja, performa tinggi ini diikuti pula dengan
skalabilitas yang tinggi, yang berarti dapat disesuaikan dengan
tingkat kebutuhan dan masalah yang dihadapi.

Penelitian ini dilakukan untuk mengetahui tingkat efisiensi dan
efektifitas penerapan sistem arsitektur cluster terhadap aplikasi
berat semacam NAMD. Penelitian ini menggunakan Linux sebagai sisem
operasi pada tiap clusternya untuk mengimplementasikan pustaka
Charm++ pada NAMD kemudian melakukan pengukuran kinerja untuk
mengetahui sejauh mana tingkat efisiensi dan efektifitas NAMD
diatas sistem cluster.

Metode yang digunakan dalam penelitian ini adalah dengan melakukan
studi literatur, pengujian dan kemudian melakukan analisa hasil
pengujian dan menarik kesimpulan. Pembahasan penelitian ini
meliput pendahuluan, landasan teori, pengujian dan analisa dan
terakhir penutup

\section{Landasan Teori}

\subsection{Arsitektur Cluster}

\textbf{Cluster} merupakan sebuah terminologi umum yang berarti:
kumpulan komputer-komputer independen dalam sebuah unifikasi
sistem lewat software dan jaringan. Pada dasarnya, dua atau lebih
komputer yang digunakan untuk memecahkan sebuah masalah
bersama-sama, dapat dikategorikan sebagai cluster.

Cluster biasanya digunakan untuk tujuan HA (High Availability)
atau HPC (High Performance Computing). Tipe yang pertama lebih
ditujukan kepada reliabilitas tinggi atau kestabilan dari sebuah
sistem. Sedangakan tipe yang kedua menjanjikan tenaga komputasi
yang jauh lebih besar dibandingkan tenaga komputer uniprosesor.

Cluster HPC sering disebut sebagai cluster \textbf{Beowulf}.
Cluster jenis ini merupakan sistem dengan performa dan
skalabilitas tinggi, menggunakan infrastruktur jaringan
\textit{private} dan sistem operasi \textit{open-source} seperti
Linux. Kinerja dapat ditingkatkan dengan menambahkan mesin kedalam
suatu sistem. Hardware mesin yang digunakan sangat bervariasi,
sebanyak yang dapat ditemukan di pasaran, mulai dari 2 (dua) node
PC \textit{stand-alone} dengan Linux dan pemakaian \textit{file
system} bersama, sampai 1024 node di atas jaringan
\textit{low-latency,} berkecepatan sangat tinggi. Cluster terbagi
kedalam 2 (dua) kelas, yaitu:

\begin{itemize}
    \item {\footnotesize Class I Cluster yang dibangun menggunakan
hardware/software umum yang ada dipasaran, dengan  teknologi standar
seperti IDE, SCSI, dan Ethernet.}
    \item {\footnotesize Class II  Cluster berbiaya tinggi, yang dibangun
menggunakan hardware khusus berkecepatan tinggi, untuk mencapai tingkat
performa terbaik.}
\end{itemize}

\subsection{Aplikasi Simulasi Dinamika Molekular NAMD}

Proses simulasi dinamika molekular menghitung kedudukan atom
dengan memecahkan persamaan-persamaan dari pergerakannya secara
numerik. Proses perhitungan ini dibantu oleh rumus medan energi
empiris yang memperkirakan energi atom dalam sistem biopolimer
secara aktual. Awalnya, program simulasi dinamika molekular
dibangun diatas mesin-mesin serial. Namun untuk menghitung proses
simulasi molekular yang lebih besar, mutlak diperlukan kekuatan
komputasi yang juga lebih besar. Salah satu cara untuk dapat
menjalankan simulasi semacam ini adalah dengan menggunakan
komputer paralel. Selain faktor kecepatan, komputer paralel juga
lebih unggul dari segi biaya.

NAMD merupakan program paralel pada UNIX yang dirancang khusus
untuk simulasi  dinamika molekular struktur biologi. NAMD
dirancang untuk berjalan dengan efisien diatas mesin-mesin
paralel. Software NAMD merupakan properti intelektual dari The
Board of Trustees of theUniversity of Illinois, mengatasnamakan
The Theoretical Biophysics Group pada Beckman Institute.Pada saat
penulisan ini dibuat, NAMD telah mencapai versi rilis 2.5, yang
juga digunakan dalam pengujian ini nantinya.

\subsection{Dasar-dasar Simulasi NAMD}
Untuk menjalankan sebuah simulasi, NAMD memerlukan empat macam
file masukan, yaitu:

\begin{enumerate}
    \item {\footnotesize File PDB (Protein Data Bank)}
    \item {\footnotesize File PSF (Protein Structure File)}
    \item {\footnotesize File parameter medan energi (force field parameter) }
    \item {\footnotesize File konfigurasi NAMD}
\end{enumerate}

File PDB menyimpan data koordinat atom dan atau kecepatan dari
suatu sistem molekular.  File ini menyimpan keseluruhan informasi
mengenai nama, jenis, dan juga jaringan molekul tersebut. Lebih
lengkapnya, dalam file ini tersimpan data pengarang, catatan
revisi, catatan jurnal, referensi, sekuen asam amino,
\textit{stoichiometry}, lokasi struktur sekunder, \textit{cyrstal
lattice}, kelompok simetri, serta catatan ATOM dan HET-ATM.
Catatan ATOM dan HET-ATM inilah yang menyimpan koordinat-koordinat
dari protein-protein, air-air, ion-ion, dan macam-macam atom
heterogen kristal lainnya.

Arsitektur file PDB memungkinkan penyimpanan lebih dari satu set
koordinat atom-atom.File PDB digunakan sebagai format data masukan
dan keluaran pada simulasi NAMD.  File PSF menyimpan informasi
struktural dari suatu protein secara spesifik. Informasi ini
terbagi atas lima bagian utama yaitu mengenai macam-macam
atom,\hspace{15pt} \textit{bond}, \textit{angle},
\textit{dihedral}, dan \textit{improper}. Kesemuanya diperlukan
untuk pengaplikasian medan energi tertentu pada simulasi nantinya.

File parameter ini berisikan konstanta numerik yang diperlukan
untuk mengevaluasi gaya dan energi pada struktur dan koordinat
atom-atom (disediakan oleh file PSF dan PDB). Parameter ini
berguna antara lain untuk mengatur panjang \textit{equilibrium}
dan kekuatan ikatan antar atom.File konfigurasi NAMD berisikan
semua konfigurasi dan pilihan yang dibutuhkan NAMD untuk
menjalankan sebuah simulasi. Dengan kata lain, file ini
memberitahu NAMD bagaimana sebuah simulasi harus dijalankan.
Sebuah simulasi dinamika molekular membutuhkan kontrol masukan
seperti temperatur, langkah waktu (time step), lama simulasi,
pengaktifan fasilitas tertentu, nama file input, nama file output,
dan berbagai parameter spesifik lainnya.

\subsection{Charm++ Sebagai Pustaka Paralel Pada NAMD}

Charm++ merupakan sebuah \textit{runtime library} yang mengizinkan
komunikasi antar objek- objek C++ dengan sangat efisien. Charm++
lebih merupakan ekstensi paralel untuk bahasa C++, yang
dikembangkan oleh PPL (Parallel Programming Laboratory) dalam
beberapa tahun terakhir. Charm++  menggunakan model pemrograman
SPMD (Single Program Multiple Data) yang dipopulerkan oleh MPI.
Model pemrograman ini bisa dibilang sangat mirip dengan CORBA,
Java RMI, atau RPC, namun lebih difokuskan kepada mesin-mesin
paralel kinerja tinggi.

\subsection{Aplikasi VMD}

VMD (Visual Molecular Dynamics) merupakan aplikasi yang dirancang
untuk memvisualisasikan dan menganalisa sistem biopolimer
(protein, asam nukleid, lipid, dan membran. VMD berjalan diatas
mayoritas sistem UNIX, Apple MacOS X, dan Microsoft Windows.
Seperti halnya NAMD, VMD juga dikembangakan oleh Theoritical
Biophysics Group, di University of Illinois and Beckman Institute,
Urbana. VMD memang dikembangkan sebagai aplikasi visual bagi
aplikasi simulasi dinamika molekular NAMD

\section{Pengujian dan Analisa}

\subsection{Konfigurasi Mesin Cluster}
Keseluruhan pengujian yang dilakukan pada penulisan ini dijalankan
diatas mesin cluster yang sama. Mesin cluster tempat pengujian ini
dijalankan terdiri dari 5 (lima) buah PC, 1 (satu) buah monitor,
dan 1 (satu) buah ethernet switch. Perangkat pendukung lainnya
seperti keyboard dan mouse hanya digunakan pada node server saja.
Kehandalan Linux sebagai sistem operasi jaringan memastikan
pengontrolan node dapat dilakukan melalui server dengan sangat
mudah.

\subsection{Metode Pengujian}
Pengujian dilakukan sebanyak 3 (tiga) kali untuk masing-masing
simulasi dan atau konfigurasi mesin yang berbeda. Tiap-tiap
simulasi dilakukan dalam kondisi \textit{fresh} setelah
me-\textit{restart }mesin terlebih dahulu. Setelah 3 kali
pengujian, mesin akan di-restart kembali untuk memulihkan kondisi
memori sebelum simulasi berikutnya dilakukan. Nilai rata-rata
kemudian diambil untuk diolah lebih lanjut sebagai perbandingan
dengan nilai hasil simulasi lainnya.

Pengujian ini sendiri dilakukan dengan berbagai metode, dalam
kaitannya pada pembuktian efektifitas dan efisiensi NAMD diatas
mesin cluster. Sebagai tambahan, akan dilakukan semacam simulasi
\textit{benchmark} untuk mencoba mengukur kinerja maksimal dan
batas ketahanan mesin cluster.

\subsection{Pengukuran Speedup Mesin Cluster}

Tidak ada cara yang lebih tepat untuk menilai efektifitas suatu
mesin paralel selain dengan mengukur  \textit{speedup } sistem
tersebut. Nilai ini didapat dengan membandingkan hasil kerja
sekuensial dengan hasil paralel untuk menilai tingkat percepatan
yang dicapai mesin paralel.

Simulasi pengujian ini dilakukan dengan menggunakan data molekular
ER-GRE (36.573 atom, spherical). Simulasi ini terdiri dari 500
numsteps, dengan temperatur sebesar 300 $^o$ Kelvin, dan
menggunakan sampai empat file parameter medan energi. Hasil
simulasi pengujian dengan penggunaan jumlah mesin yang berbeda
dapat dilihat pada tabel berikut.

\begin{table}
\centering
\begin{tabular}{|p{64pt}|p{70pt}|p{70pt}|p{71pt}|}
\hline
\parbox{64pt}{\centering
Banyak Node } & \parbox{70pt}{\centering Rata-rata WallClock (s) }
& \parbox{70pt}{\centering Rata-Rata CPUTime (s) } &
\parbox{71pt}{\centering Rata-rata Memory Used (kB)
} \\
\hline
\parbox{64pt}{\centering
1
} & \parbox{70pt}{\raggedleft
989,563009
} & \parbox{70pt}{\raggedleft
984,696676
} & \parbox{71pt}{\raggedleft
83.439
} \\
\parbox{64pt}{\centering
2
} & \parbox{70pt}{\raggedleft
539,414144
} & \parbox{70pt}{\raggedleft
529,566671
} & \parbox{71pt}{\raggedleft
17.571
} \\
\parbox{64pt}{\centering
3
} & \parbox{70pt}{\raggedleft
402,206553
} & \parbox{70pt}{\raggedleft
368,890005
} & \parbox{71pt}{\raggedleft
15.694
} \\
\parbox{64pt}{\centering
4
} & \parbox{70pt}{\raggedleft
345,493255
} & \parbox{70pt}{\raggedleft
297,353333
} & \parbox{71pt}{\raggedleft
16.033
} \\
\parbox{64pt}{\centering
5
} & \parbox{70pt}{\raggedleft
259,569122
} & \parbox{70pt}{\raggedleft
234,509999
} & \parbox{71pt}{\raggedleft
15.956
} \\
\hline
\end{tabular}
\caption{Nilai rata-rata lama waktu simulasi ER-GRE.}\label{T1}
\end{table}

\begin{figure}
  \includegraphics[width=351pt]{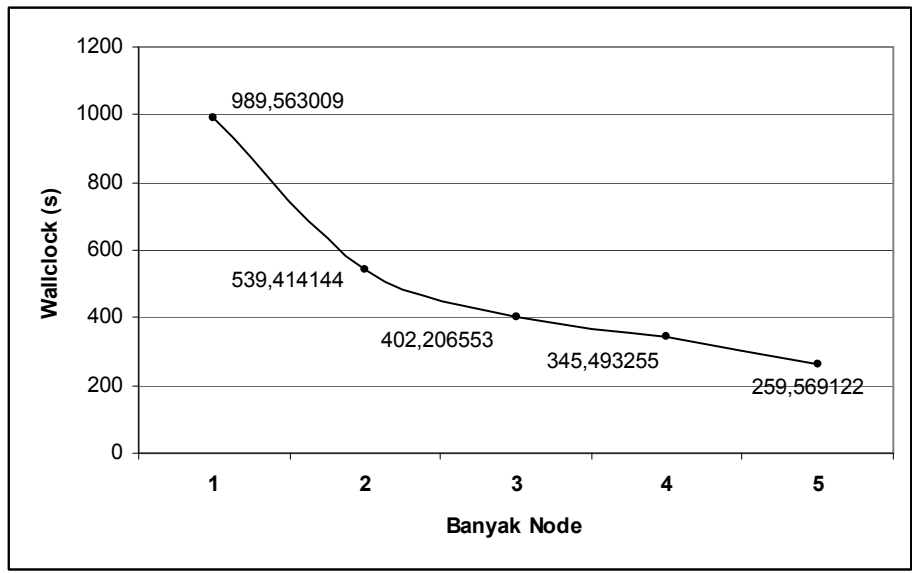}
\\
  \caption{Grafik nilai rata-rata lama waktu
simulasi ER-GRE.}\label{g1}
\end{figure}

Menggunakan hasil diatas, kemudian dapat dihitung nilai speedup
mesin cluster. Speedup diperoleh dengan membagi waktu sekuensial
dengan waktu paralel tiap-tiap pengujian. Persamaannya sebagai
berikut:

\begin{center}

{\footnotesize Speed up = waktu sekuensial/waktu paralel = Ts/Tp
          (1)}

\end{center}

\begin{table}
  \centering
\begin{tabular}{|p{63pt}|p{92pt}|p{92pt}|p{92pt}|}
\hline
\parbox{63pt}{\centering
Banyak Node } & \parbox{92pt}{\centering Rata-rata WallClock (s) }
& \parbox{92pt}{\centering Speedup } &
\parbox{92pt}{\centering Speedup Ideal
} \\
\hline
\parbox{63pt}{\centering
1
} & \parbox{92pt}{\raggedleft
989,563009
} & \parbox{92pt}{\centering
-
} & \parbox{92pt}{\centering
-
} \\
\parbox{63pt}{\centering
2
} & \parbox{92pt}{\raggedleft
539,414144
} & \parbox{92pt}{\raggedleft
1,83
} & \parbox{92pt}{\raggedleft
2 kali
} \\
\parbox{63pt}{\centering
3
} & \parbox{92pt}{\raggedleft
402,206553
} & \parbox{92pt}{\raggedleft
2,46
} & \parbox{92pt}{\raggedleft
3 kali
} \\
\parbox{63pt}{\centering
4
} & \parbox{92pt}{\raggedleft
345,493255
} & \parbox{92pt}{\raggedleft
2,86
} & \parbox{92pt}{\raggedleft
4 kali
} \\
\parbox{63pt}{\centering
5
} & \parbox{92pt}{\raggedleft
259,569122
} & \parbox{92pt}{\raggedleft
3,81
} & \parbox{92pt}{\raggedleft
5 kali
} \\
\hline
\end{tabular}

\caption{Nilai speedup simulasi ER-GRE pada mesin
cluster.}\label{T2}
\end{table}

Dari angka-angka diatas dapat terlihat bahwa tingkat percepatan
yang dicapai oleh sistem cluster ini masih jauh dari ideal. Jika
dilihat pada pemakaian 2 buah prosesor (2 node), speedup yang
diperoleh mencapai tingkat 1,83 kali lebih cepat. Peningkatan ini
sangat berarti karena dapat menghemat waktu sampai 450 detik. Pada
pemakaian 3 buah node peningkatan ini semakin menurun, dan pada
pemakaian 4 node, speedup yang dicapai bahkan masih belum
melampaui nilai 3.

\begin{figure}
  \includegraphics[width=351pt]{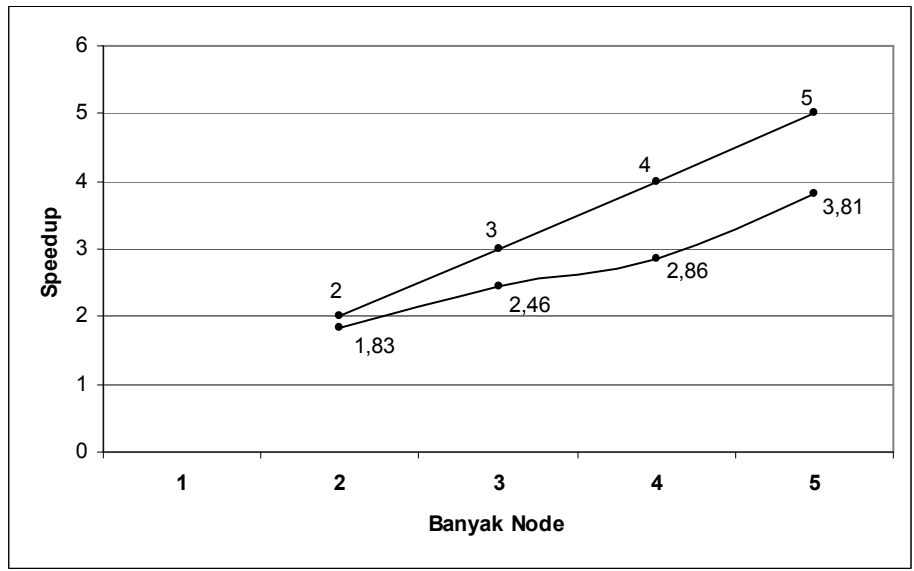}\\
  \caption{Grafik speedup simulasi ER-GRE pada
mesin cluster.}\label{g2}
\end{figure}

\subsection{Tingkat Efisiensi Paralel Mesin Cluster}

Pengujian tingkat efisiensi sebuah mesin paralel merupakan sebuah
tolak ukur yang bersifat objektif. Besar objek atau data simulasi
yang menjadi bahan pengujian akan menentukan tingkat efisiensi
sebuah mesin paralel. Oleh karena itu, untuk mesin dan besar data
yang berbeda, akan diperoleh tingkat efisiensi yang berbeda pula.

Pengujian kali ini dilakukan masih dengan menggunakan data
molekular ER-GRE (36.573 atom, spherical). Oleh karena itu hasil
pengujian sebelumnya masih bisa digunakan untuk mencari efisiensi
paralel mesin cluster. Efisiensi paralel didefinisikan dengan
persamaan berikut:

\begin{center}

{\footnotesize Efisiensi Paralel (\begin{math}\eta{}\end{math}) =
100\% x waktu sekuensial/(P x waktu Paralel)         (2)}
\end{center}

\begin{table}
  \centering

\begin{tabular}{|p{64pt}|p{91pt}|p{91pt}|p{91pt}|}
\hline
\parbox{64pt}{\centering
Banyak Node } & \parbox{91pt}{\centering Rata-rata WallClock (s) }
& \parbox{91pt}{\centering Jumlah Pemroses x Waktu Paralel (s) } &
\parbox{91pt}{\centering Efisiensi Paralel (\%)
} \\
\hline
\parbox{64pt}{\centering
1
} & \parbox{91pt}{\raggedleft
989,563009
} & \parbox{91pt}{\centering
-
} & \parbox{91pt}{\centering
-
} \\
\parbox{64pt}{\centering
2
} & \parbox{91pt}{\raggedleft
539,414144
} & \parbox{91pt}{\raggedleft
1.078,828288
} & \parbox{91pt}{\raggedleft
91,73\%
} \\
\parbox{64pt}{\centering
3
} & \parbox{91pt}{\raggedleft
402,206553
} & \parbox{91pt}{\raggedleft
1.206,619659
} & \parbox{91pt}{\raggedleft
82,01\%
} \\
\parbox{64pt}{\centering
4
} & \parbox{91pt}{\raggedleft
345,493255
} & \parbox{91pt}{\raggedleft
1.381,973020
} & \parbox{91pt}{\raggedleft
71,61\%
} \\
\parbox{64pt}{\centering
5
} & \parbox{91pt}{\raggedleft
259,569122
} & \parbox{91pt}{\raggedleft
1.297,845610
} & \parbox{91pt}{\raggedleft
76,25\%
} \\
\hline
\end{tabular}

\caption{Tingkat efisiensi paralel simulasi ER-GRE pada mesin
cluster.}\label{T3}
\end{table}

Tabel di atas memperlihatkan bahwa tingkat efisiensi paralel mesin
cluster dalam menjalankan simulasi ER-GRE sangat memuaskan. Meski
terlihat penurunan tingkat efisiensi pada setiap penambahan jumlah
node, penurunan ini masih dalam batas yang wajar. Ada fenomena
yang sedikit menarik perhatian pada pengujian ini. Pada pemakaian
5 buah node, tingkat efisiensi terlihat mengalami kenaikan dari
71,61\% menjadi 76,25\%. Kira-kira apa yang akan terjadi apabila
dilakukan pengujian pada 6 buah node? Ada kemungkinan, tingkat
efisiensi paralel ini akan mengalami kenaikan lagi. Walaupun hal
ini tidak dapat dibuktikan, karena keterbatasan mesin pengujian
dan lain hal, fakta ini tetap menunjukkan satu hal: bahwa dalam
simulasi tersebut, pemakaian 5 mesin masih lebih efisien
dibandingkan pemakaian 4 mesin. Untuk saat ini, fenomena tersebut
hanya dapat dianggap sebagai faktor acak (random factor) yang
tidak terduga. Faktor acak ini sangat erat kaitannya dengan apa
yang disebut dengan ongkos komunikasi atau \textit{communication
cost}.

\subsection{Ongkos Komunikasi}

Pada pemecahan suatu masalah dengan cara paralel, biasanya tingkat
efisiensi akan menurun saat jumlah pemroses dinaikkan. Kenapa hal
ini bisa terjadi? Hal ini disebabkan tidak lain oleh terjadinya
ongkos komunikasi. Ongkos komunikasi merupakan faktor yang sangat
dominan terjadi pada proses komputasi paralel. Ongkos komunikasi
adalah lama waktu yang digunakan untuk proses pembuatan,
pengiriman, dan penerimaan data. Semakin banyak jumlah pemroses,
dalam hal ini jumlah node yang digunakan, maka akan semakin besar
nilai ongkos komunikasi tersebut. Angka ini akan menjadi semakin
besar jika tingkat masalah yang harus dipecahkan semakin kecil,
karena waktu yang dibutuhkan untuk memecahkan masalah malah lebih
singkat ketimbang waktu komunikasi data itu sendiri.  Sebagai
pembuktian, akan dilakukan simulasi pengujian dengan menggunakan
data molekular decalanin (66 atom, tiny). Simulasi ini terdiri
dari 1000 numsteps, dengan temperatur sebesar 300$^o$ K. Lebih
jelasnya dapat dilihat pada file konfigurasi untuk simulasi
decalanin. Pada tabel berikut, dapat dilihat lama waktu simulasi
pengujian pada mesin cluster:

\begin{table}
  \centering

\begin{tabular}{|p{64pt}|p{91pt}|p{91pt}|p{91pt}|}
\hline
\parbox{64pt}{\centering
Banyak Node } & \parbox{91pt}{\centering WallClock (s) } &
\parbox{91pt}{\centering Jumlah Pemroses x Waktu Paralel
(s) } & \parbox{91pt}{\centering Efisiensi Paralel (\%)
} \\
\hline
\parbox{64pt}{\centering
1
} & \parbox{91pt}{\raggedleft
6,792329
} & \parbox{91pt}{\centering
-
} & \parbox{91pt}{\centering
-
} \\
\parbox{64pt}{\centering
2
} & \parbox{91pt}{\raggedleft
17,772844
} & \parbox{91pt}{\raggedleft
35,545688
} & \parbox{91pt}{\raggedleft
19,11\%
} \\
\parbox{64pt}{\centering
3
} & \parbox{91pt}{\raggedleft
18,340601
} & \parbox{91pt}{\raggedleft
55,021803
} & \parbox{91pt}{\raggedleft
12,34\%
} \\
\hline
\end{tabular}
\caption{Tingkat efisiensi paralel simulasi decalanin pada mesin
cluster.}\label{T4}
\end{table}

Pada tabel diatas dapat dilihat bahwa lama waktu simulasi semakin
bertambah seiring kenaikan jumlah pemroses. Kasus ini menjadi
semacam antiklimaks bagi komputasi paralel. Simulasi ini dapat
menunjukkan secara sempurna, untuk tingkat permasalahan yang
relatif kecil, sangat tidak  efisien jika diselesaikan secara
paralel. Waktu komunikasi disini terbukti jauh lebih lama
dibandingkan waktu pemrosesan itu sendiri. Jika diasumsikan bahwa
tingkat efisiensi paralel adalah ideal, akan diperoleh suatu
persamaan:

\begin{center}

{\footnotesize \begin{math}\eta{}\end{math} = Ts/(P x Tp) = 100 \%}

\end{center}

\begin{center}

{\footnotesize                                    Tp= Ts/P
                                           (3)}

\end{center}

Menyertakan faktor ongkos komunikasi, maka

\begin{center}
{\footnotesize Tp = Ts/P +  Tcomm}
\end{center}
\begin{center}
{\footnotesize Tp = Ts (1 +y) /P (4)}
\end{center}

Dimana y adalah rasio komunikasi dengan komputasi. Fakta ini
lagi-lagi memberikan satu analogi, bahwa apabila tingkat
permasalahandiperbesar, maka nilai y (rasio komunikasi dengan
komputasi) akan berkurang, yang berarti tingkat efisiensi paralel
menjadi meningkat.

Suatu algoritma paralel dikatakan memiliki skalabilitas secara
isoefisien , jika saat jumlah pemroses (P) dinaikkan, tingkat
efisiensi paralel tetap dapat dijaga dengan memperbesar ukuran
masalah yang akan dipecahkan. Misalkan, sebuah algoritma paralel
mampu melakukan simulasi sejumlah atom (N1) pada beberapa pemroses
(P1), dengan efisiensi sebesar
\begin{math}\eta{}\end{math}. Maka, saat jumlah pemroses diperbesar
(P2),efisiensi (\begin{math}\eta{}\end{math}) tetap dapat
dipertahankan dengan menaikkan besar masalah sedemikian rupa (N2).
Hal ini dapat ditunjukkan dengan persamaan:

\begin{center}
{\footnotesize Ts1/(P1 (1+y1) Ts1/P1) = Ts2/(P2 (1+y2) Ts2/P2)}
\end{center}
\begin{center}
{\footnotesize 1/(1+y1) = 1/(1+y2)}
\end{center}
\begin{center}
{\footnotesize y1 = y2 (5)}
\end{center}

Kesimpulannya, tingkat efisiensi tetap dapat dijaga dengan
memperbesar permasalahan sedemikian rupa, sehingga nilai y (rasio
komunikasi dengan komputasi) tetap sama.

\subsection{Simulasi Benchmark}

Pengujian berikut ini lebih bertujuan untuk mengetahui batas
kemampuan mesin cluster dalam menjalankan simulasi skala besar.
Untuk simulasi pengujian kali ini, data molekular yang akan
dipakai adalah ApoA1 (92.224 atom, periodic). Simulasi ini tidak
akan dijalankan secara paralel bertahap, melainkan langsung
memakai seluruh kemampuan mesin cluster yang ada (5 mesin).
Sebagai catatan, simulasi ApoA1 yang hanya dijalankan diatas 1
mesin (cluster1) saja, memakan waktu kurang lebih 4 hari (99,88403
jam). Waktu ini lebih berupa harga estimasi, karena simulasi tadi
sebenarnya tidak pernah terselesaikan, karena keterbatasan waktu
dan lain hal. Kemungkinan besar, akan dibutuhkan waktu yang lebih
lama lagi untuk menyelesaikan simulasi tadi secara benar. Simulasi
ini terhitung berat karena menyertakan perhitungan elektrostatis
penuh, PME (Particle Mesh Ewald), ditambah penggunaan 2 (dua) file
parameter medan energi. Simulasi pengujian akan dilakukan sebanyak
11 (sebelas) kali, masing-masing dengan besar numsteps yang
berbeda. Hasil pengujian simulasi pada mesin cluster dapat dilihat
pada tabel berikut

\begin{table}
  \centering

\begin{tabular}{|p{65pt}|p{91pt}|p{91pt}|p{91pt}|}
\hline
\parbox{65pt}{\centering
Jumlah \textit{numsteps} } &
\parbox{91pt}{\centering Rata-rata WallClock
(s) } & \parbox{91pt}{\centering Rata-Rata CPUTime (s) } &
\parbox{91pt}{\centering Rata-rata Memory
Used (kB)
} \\
\hline
\parbox{65pt}{\centering
{\scriptsize 500}
} & \parbox{91pt}{\raggedleft
{\scriptsize 880,857320}
} & \parbox{91pt}{\raggedleft
{\scriptsize 846,403341}
} & \parbox{91pt}{\raggedleft
{\scriptsize 43.886}
} \\
\parbox{65pt}{\centering
{\scriptsize 600}
} & \parbox{91pt}{\raggedleft
{\scriptsize 1.016,335510}
} & \parbox{91pt}{\raggedleft
{\scriptsize 1.001,873352}
} & \parbox{91pt}{\raggedleft
{\scriptsize 42.570}
} \\
\parbox{65pt}{\centering
{\scriptsize 700}
} & \parbox{91pt}{\raggedleft
{\scriptsize 1.173,281861}
} & \parbox{91pt}{\raggedleft
{\scriptsize 1.156,503337}
} & \parbox{91pt}{\raggedleft
{\scriptsize 42.571}
} \\
\parbox{65pt}{\centering
{\scriptsize 800}
} & \parbox{91pt}{\raggedleft
{\scriptsize 1.328,010783}
} & \parbox{91pt}{\raggedleft
{\scriptsize 1.309,586670}
} & \parbox{91pt}{\raggedleft
{\scriptsize 42.571}
} \\
\parbox{65pt}{\centering
{\scriptsize 900}
} & \parbox{91pt}{\raggedleft
{\scriptsize 1.486,034872}
} & \parbox{91pt}{\raggedleft
{\scriptsize 1.463,863363}
} & \parbox{91pt}{\raggedleft
{\scriptsize 42.572}
} \\
\parbox{65pt}{\centering
{\scriptsize 1000}
} & \parbox{91pt}{\raggedleft
{\scriptsize 1.646,129720}
} & \parbox{91pt}{\raggedleft
{\scriptsize 1.622,039998}
} & \parbox{91pt}{\raggedleft
{\scriptsize 42.571}
} \\
\parbox{65pt}{\centering
{\scriptsize 1100}
} & \parbox{91pt}{\raggedleft
{\scriptsize 1.809,191447}
} & \parbox{91pt}{\raggedleft
{\scriptsize 1.763,896688}
} & \parbox{91pt}{\raggedleft
{\scriptsize 42.569}
} \\
\parbox{65pt}{\centering
{\scriptsize 1200}
} & \parbox{91pt}{\raggedleft
{\scriptsize 1.963,564820}
} & \parbox{91pt}{\raggedleft
{\scriptsize 1.924,103312}
} & \parbox{91pt}{\raggedleft
{\scriptsize 42.571}
} \\
\parbox{65pt}{\centering
{\scriptsize 1300}
} & \parbox{91pt}{\raggedleft
{\scriptsize 2.124,480713}
} & \parbox{91pt}{\raggedleft
{\scriptsize 2.077,383301}
} & \parbox{91pt}{\raggedleft
{\scriptsize 42.571}
} \\
\parbox{65pt}{\centering
{\scriptsize 1400}
} & \parbox{91pt}{\raggedleft
{\scriptsize 2.286,481689}
} & \parbox{91pt}{\raggedleft
{\scriptsize 2.232,583333}
} & \parbox{91pt}{\raggedleft
{\scriptsize 42.571}
} \\
\parbox{65pt}{\centering
{\scriptsize 1500}
} & \parbox{91pt}{\raggedleft
{\scriptsize 2.431,499756}
} & \parbox{91pt}{\raggedleft
{\scriptsize 2.383,686605}
} & \parbox{91pt}{\raggedleft
{\scriptsize 42.571}
} \\
\hline
\end{tabular}
\caption{Nilai rata-rata lama waktu simulasi ApoA1.}\label{T5}
\end{table}

\begin{figure}
  \includegraphics[width=351pt]{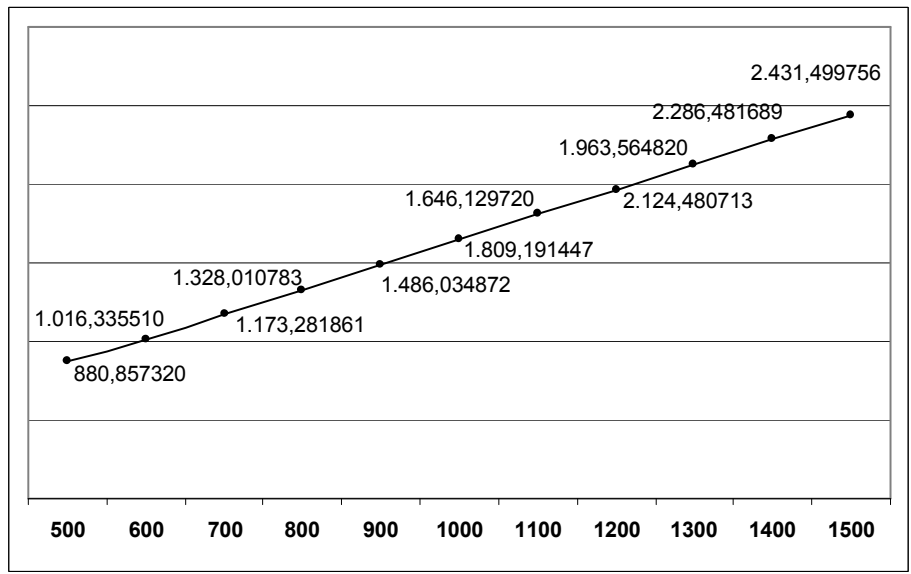}\\
  \caption{Grafik nilai rata-rata lama waktu
simulasi ApoA1.}\label{g3}
\end{figure}

\section{Penutup}

Melalui hasil pengujian dan analisa yang diperoleh sebelumnya,
dapat diambil kesimpulan bahwa pengaturan konfigurasi hardware
memiliki kontribusi besar dalam kinerja keseluruhan mesin cluster.
Terbukti bahwa kesalahan kecil dalam pengaturan BIOS, misalnya
besar \textit{shared-memory }untuk VGA, dapat membawa perubahan
besar dalam hal kecepatan kerja.

Pemilihan distribusi Linux yang tepat (SuSE 9.0) ditambah dengan
penggunaan kernel vanilla terbaru (2.4.21) dapat sedikit menambah
kecepatan kerja paralel mesin. Penambahan kecepatan ini mungkin
tidak terlalu signifikan, akan tetapi dalam menjalankan simulasi
skala besar, perbedaan ini akan terasa sekali.

Ongkos komunikasi (communication cost) yang lebih besar
dibandingkan waktu pemrosesan dapat membatasi tingkat efisiensi
suatu komputasi paralel. Faktor-faktor penyebab besarnya nilai
ongkos komunikasi ini antara lain: delay komunikasi, delay
sinkronisasi, ketidakseimbangan beban, waktu pembuatan proses,
kode sekuensial yang berlebihan dan masalah memory contention

NAMD terbukti memiliki algoritma paralel dengan skalabilitas
tinggi, karena saat jumlah pemroses dinaikkan, tingkat efisiensi
tetap dapat dijaga dengan memperbesar ukuran masalah. Mesin
cluster pengujian dapat dikatakan memiliki tingkat efisiensi dan
efektifitas yang cukup tinggi. Terbukti dengan banyaknya waktu
yang dapat dihemat saat menjalankan simulasi-simulasi besar.
Penurunan tingkat efisiensi lebih disebabkan oleh ukuran
permasalahan yang tidak lagi cukup besar untuk layak diproses pada
banyak node.

Binari NAMD yang dikompilasi secara lengkap (menyertakan TCL,
FFTW, VMD plugins molfile) akan menjalankan simulasi-simulasi
tertentu dengan lebih cepat. Pada penyelesaian simulasi  ApoA1
yang menyertakan perhitungan PME (Particle Mesh Ewald), waktu yang
dapat dihemat cukup signifikan.

Mesin cluster terbukti pula mampu menjalankan simulasi interaktif
secara \textit{real-time} tanpa masalah berarti. IMD (Interactive
Molecular Dynamics) akan menjadi standar teknologi dalam dunia
simulasi dinamika molekular di masa mendatang.

Melalui berbagai pengalaman dan pelajaran menarik yang dialami,
ada beberapa saran yang untuk mengembangkan penelitian ini lebih
jauh. Pada penelitian ini, kemampuan grafis mesin cluster
dirasakan kurang memadai untuk menjalankan simulasi IMD dengan
jumlah atom yang lebih besar. Dibutuhkan lebih dari sekedar tenaga
prosesor saja untuk menjalankan simulasi interaktif secara
\textit{real-time} sekaligus memvisualisasikannya.

Akan sangat menarik bila arsitektur sistem cluster ini dapat
dicobakan pada mesin-mesin yang lebih cepat, yang dilengkapi
dengan teknologi infrastruktur jaringan terbaru. Mesin cluster
yang dibangun di atas jaringan gigabit ethernet (1000 Mbps)
merupakan impian setiap pemrogram dan praktisi teknologi
informasi, khususnya yang berkecimpung dalam bidang komputasi
paralel.

\end{document}